\documentclass[10pt,preprint]{aastex}

\newcommand{\sersic}{S\'{e}rsic}

\begin{document}

\title{Imaging of Three Possible Low Redshift Analogs to High Redshift Compact Red Galaxies}
\author{Hsin-Yi Shih and Alan Stockton\affil{Institute for Astronomy, 
University of Hawaii, 2680 Woodlawn Dr, Honolulu, HI 96822  
{\it hsshih@ifa.hawaii.edu, stockton@ifa.hawaii.edu}}}

\begin{abstract}

As part of a larger program to identify and characterize possible low-redshift analogs to massive compact red galaxies found at high redshift, we have examined the morphologies of three low-redshift compact galaxies drawn from the sample of \citet{trujillo09}. Using deeper and higher resolution images, we have found faint and relatively extensive outer structures in addition to the compact cores identified in the earlier measurements. One object appears to have a small companion that may be involved in an ongoing minor merger of the sort that could be responsible for building up the outer parts of these galaxies. The ages of the dominant stellar populations in these objects are found to be around 2--4 Gyr, in good agreement with the previous estimates. The presence of diffuse outer structures in these galaxies indicates that truly compact and massive red galaxies are exceedingly rare at low redshift. The relatively young stellar populations suggest that the accretion of the extensive outer material must occur essentially universally on relatively short timescales of few billion years or less. These results confirm and extend previous suggestions that the driving mechanism behind the size evolution of high redshift compact galaxies cannot be highly stochastic processes such as major mergers, which would inevitably leave a non-negligible fraction of survivors at low redshift. 

\end{abstract}

\maketitle

\section{Introduction}

Morphologies of high redshift galaxies are important in providing constraints on galaxy formation models. High-redshift massive galaxies showing little or no recent star formation have been found to be more compact than their lower-redshift counterparts of similar masses \citep*[e.g.,][and references therein]{stockton04, daddi05, trujillo07, zirm07, toft07, dokkum08, franx08, buitrago08, damjanov09, muzzin09} and many recent efforts have been made to understand their nature and subsequent evolution \citep*[e.g.,][]{cimatti08, hopkins09, dokkum10, wuyts10}. \citet{kriek06} have found that $\sim45$ \% of the massive $K$-band-selected galaxies at $z \sim 2.3$ have old stellar populations. Van Dokkum et al.\ (2008), using high resolution images of the \citeauthor{kriek06} sample from HST and laser-guide-star adaptive-optics, estimated that $\sim 90\!-\!100$\% of these massive galaxies with old stellar populations are extremely compact, with mean effective radii of $< 1$ kpc.

Such galaxies present an evolutionary puzzle, since they are not present or at least are exceedingly rare in the present-day universe. It has been suggested that these high-redshift compact galaxies might have evolved into present day elliptical galaxies through ``dry'' mergers \citep{khochfar06} or ``puffing up'' due to quasar feedback \citep{fan08}, among other processes. However, as pointed out by \citet{taylor10}, if the evolution involves purely stochastic processes such as mergers, some significant fraction of these galaxies is expected to survive intact and to be found in the lower-redshift universe. This possibility is of considerable interest because, at $z > 2$, these galaxies are quite faint, and since they have no emission lines, they are extremely difficult to study in any detail. As a consequence, some recent studies have focused on identifying and characterizing possible local analogs of these objects \citep*[e.g.,][]{trujillo09, taylor10, stockton10}.

\citet{trujillo09} (henceforth T09) searched the Sloan Digital Sky Survey (SDSS) DR6 NYU Value-Added Galaxy Catalog \citep*{blanton05} for nearby massive compact galaxies ($z < 0.2$, $M  > 8 \times 10^{10}M_{\odot}$,  and (circularized) $R_{e}  < 1.5$ kpc). From an original sample of 48 selected by their search criteria, they analyzed the SDSS data of 29 objects that were not impacted by nearby objects or rejected for other reasons. A search by \citet{taylor10} at a lower redshift ($z<0.12$, with almost no overlap with the range of T09) of SDSS DR7 returned no candidates as massive and as compact as those identified at high redshift. In our own earlier study, \citet{stockton10}, we found that truly small, massive, compact galaxies seem to be extremely rare even at a somewhat higher redshift. At this earlier epoch, but with a similar co-moving volume to that of T09 ($\sim 200$ deg$^{2}$ and $0.4 \le z \le 0.8$), we found just 2 confirmed cases of compact massive galaxies. While we had somewhat more stringent selection criteria (effectively, $M\gtrsim2 \times 10^{11}M_{\odot}$ and $R_{e}  < 1.0$ kpc), we still felt it important to check into the comparatively large number found by T09, particularly since the angular effective radii were comparable to, or smaller than, the SDSS pixel scale. To improve our understanding of the true morphologies of the objects in T09, we obtained deeper and higher resolution images for 3 of them using Keck I Low Resolution Imaging Spectrometer (LRIS; Oke et al. 1995). 

\section{Observations and Data Reductions}

Out of the T09 sample, we chose the 3 galaxies accessible during our April observing run with apparent masses $> 10^{11} M_\odot$, SDSS\,J090324.19+022645.3, SDSS\,J092723.34+215604.8, and SDSS\,J101637.23+390203.6. This last galaxy, SDSSJ1016, is listed by T09 to be the most compact object in their sample ($R_{e} = 0.88$ kpc). We deliberately chose galaxies from the upper end of the mass range of the T09 sample, as these are potentially the most similar to the compact passive galaxies that have so far been observed at $z>2$, which typically have masses of $\sim2\times 10^{11} M_\odot$ or more. We observed these 3 galaxies on 2010 Apr 08 UT using the red side of LRIS on Keck I telescope. The images were obtained using the $I$ filter, which is centered on 7599~\AA~and has a FWHM of 1225~\AA. These parameters are very close to those for the SDSS i filter, and we accordingly have calibrated our images on the SDSS system using unsaturated stars that have good SDSS magnitudes. A total of 3 dithered images were obtained for each object, each with an exposure time of 180 s for SDSSJ0903, and 60 s for SDSSJ0927 and SDSSJ1016. The image scale was 0\farcs135 per pixel (compared to 0\farcs396 for the SDSS), and the FWHM of the final images ranged from 0\farcs52 to 0\farcs67. The data were reduced with IRAF following standard procedures for bias subtraction and flat-fielding using dome flats, and the 3 images for each galaxy were then registered and averaged. Surface brightness limits for the final images ($3 \sigma$) were estimated from the variance of sky measurements in $4\farcs8$ apertures and amounted to $i=27.3$ for the SDSSJ0903 field and $i=26.7$ for the SDSSJ0927 and SDSSJ1016 fields. These correspond to about 2.2 and 1.6 mag fainter, respectively, than those measured from apertures of the same angular size from the SDSS $i$-band images.

\section{Data Analysis and Results}

\begin{figure*}[!t]
\epsscale{1.0}
\plotone{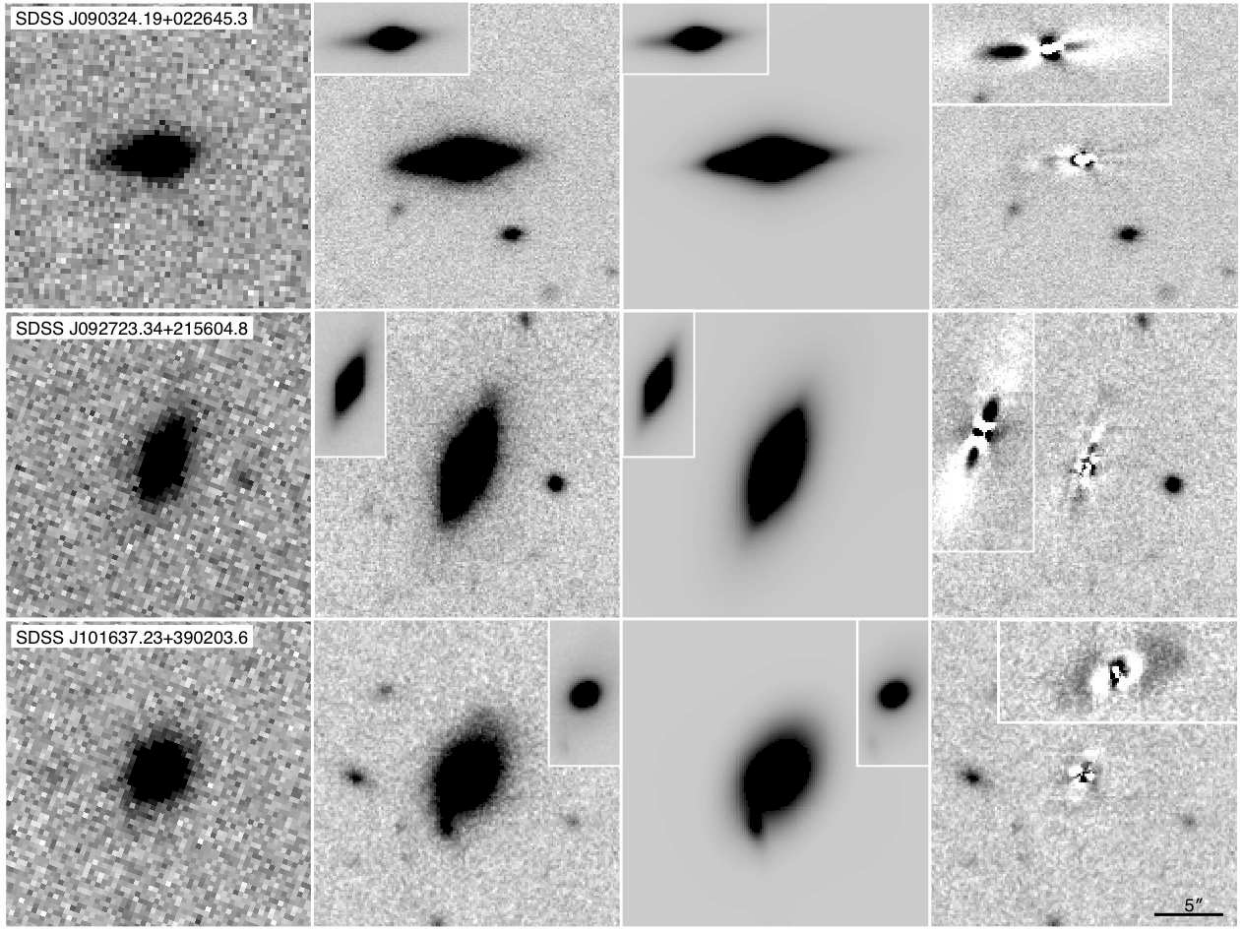}
\caption{Images of the 3 galaxies. The 1st column shows the SDSS $i$-band images; the 2nd column shows the LRIS images; the 3rd column shows the {\sc galfit} best-fit (multi-component) models, and the last column shows the residuals from subtracting the models from the images. The insets in the second and third columns show lower-contrast versions of the images and models, respectively. The insets in the last column show the residuals from the single-component fits. The image scale  for all panels is shown at the lower right, and north is up and east to the left for all images.}
\label{fits}
\end{figure*}

We used {\sc galfit} \citep*{peng02}, a 2-D galaxy profile fitting routine, to determine some of the morphological properties of our targets. PSFs were derived from unsaturated stars in the immediate fields of the galaxies by using the stellar images themselves for the inner part of the profile (down to 1--2\%\ of the peak), while fitting the wings of the stars with 2-component elliptical Moffat profiles to eliminate sky noise. Because the galaxies are resolved and turned out to be complex objects, we had to fit two or three \sersic\ components to each to get reasonably low residuals when subtracting the model from the observed image. Our goal in choosing the number of components to use was to effectively eliminate large-scale systematic residuals from the model subtractions. There was an additional problem with SDSSJ0903, in that a $\sim0\farcs4$-radius region around the center of the galaxy was either in the non-linear CCD response regime or saturated. This region was masked off for the fits for SDSSJ0903, as were all discrete objects in the rest of the field for all objects. The one exception was for SDSSJ1016, where there is a companion too close to the galaxy for effective masking. In this case, we simply included this object in the fitting procedure, using a single \sersic\ fit. The best-fit output parameters from {\sc galfit} are summarized in Table \ref{galfit3}. {\sc Galfit} outputs the effective radius in units of pixels, and the physical sizes were calculated using Ned Wright's Javascript Cosmology Calculator \citep{wright06} assuming a flat cosmology with $H_0=71$ km s$^{-1}$ Mpc$^{-1}$ and $\Omega_{M} = 0.27$ (these values differ slightly from the $H_0=70$ km s$^{-1}$ Mpc$^{-1}$, $\Omega_{M} = 0.3$ used by T09; however, scale differences between the two amount to only $\sim1$\% at the redshifts in question). While {\sc galfit} gives as the effective radius the semi-major axis value $a_e$, the values given for $R_e$ in Tables 1 and 2 are circularized (i.e., $R_e = a_e\sqrt{b/a}$, where $a_e$ is expressed in kpc). This convention ensures that half of the total light of the model is within the elliptical isophote that encloses an area of $\pi R_e^2$. The images of data, models and residuals of the best fits are shown in Fig \ref{fits}. For comparison with values in T09, we also present the output parameters of one-component fits to our LRIS images in Table \ref{galfit1}, along with our own determination of these parameters from the SDSS $i$-band images.

\begin{deluxetable}{lcccccc}
\tablecaption{{\sc Galfit} Parameters for Best Fits\label{galfit3}}
\tablewidth{0pt}
\tablecolumns{6}
\tablehead{\colhead{Object Name} & \colhead{$z$} & \colhead{$r_{e}$}  & \colhead{$R_{e}$}  & \colhead{$n$} &  \colhead{$b/a$}  & \colhead{$m_I$}\\
\colhead{} & \colhead{} & \colhead{(arcsec)} & \colhead{(kpc)} & \colhead{} & \colhead{} & \colhead{}}
\startdata
SDSS J090324.19+022645.3 & $0.187$ & 0\farcs27 & $0.84$ & $4.26$ & $0.40$ & $17.28$\\
&  & 0\farcs50 & $1.56$ & $1.29$ & $0.12$ & $18.62$\\
&  & 0\farcs67 & $2.08$ & $0.56$ & $0.30$ & $20.65$\\
\multicolumn{2}{r}{Total, all components:} & 0\farcs40 & 1.05 & & & $16.96$\\

SDSS J092723.34+215604.8 & $0.167$ & 0\farcs11 & $0.30$ & $0.80$ & $0.59$ & $18.02$\\
&  & 0\farcs46 & $1.31$ & $1.08$ & $0.14$ & $18.00$\\
&  & 1\farcs26 & $3.57$ & $1.96$ & $0.42$ & $17.89$\\
\multicolumn{2}{r}{Total, all components:} & 0\farcs37 & 1.23 & & & $16.78$\\

SDSS J101637.23+390203.6 & $0.195$ & 0\farcs30 & $0.96$ & $1.79 $ & $0.69$ & $17.78$\\
&  & 2\farcs22& $7.10$ & $0.50$ & $0.64$ & $18.54$\\
\multicolumn{2}{r}{Total, both components:} & 0\farcs55 & 1.77 & & & $17.34$\\
\enddata
\tablecomments{{\sc Galfit} output parameters of the 2 or 3 component fits for each target. Each of the lines for each object represents one Sersic component. Column 2: Object redshift. Column 3: Circularized effective radius in arcsec. Column 4: Circularized effective radius in kpc. Column 5: \sersic\ index. Column 6: Minor to major axis ratio. Column 7: Integrated magnitude of the component. For SDSSJ0903, the weak 3rd component (accounting for only 3\%\ of the total light) has been offset 3\arcsec\ to the east from the center to model the asymmetry in this object.}
\end{deluxetable}

\begin{deluxetable}{lcccccc}
\tablecaption{{\sc Galfit} Parameters for 1-Component Fits\label{galfit1}}
\tablewidth{0pt}
\tablecolumns{7}
\tablehead{\colhead{Object Name} & \colhead{$R_{e}$} & \colhead{$R_{e}$} & \colhead{$R_{e}$} & \colhead{$n$} &  \colhead{$b/a$}  & \colhead{$m_I$}\\
\colhead{} & \colhead{(LRIS)} & \colhead{(SDSS)} & \colhead{(T09)} & \colhead{} & \colhead{} & \colhead{}}

\startdata
SDSS J090324.19+022645.3 & $1.54$ & $1.02$ & $1.40$ & $1.99$ & $0.32$ & $17.11$\\
SDSS J092723.34+215604.8 & $1.26$ & $1.18$ & $1.42$ & $5.46$ & $0.26$ & $16.71$\\
SDSS J101637.23+390203.6 & $2.35$ & $1.77$ & $0.88$ & $6.39$ & $0.71$ & $17.20$
\enddata
\tablecomments{Column 2: Circularized effective radius in kpc for the LRIS images. Column 3: Circularized effective radius in kpc determined from {\sc galfit} analysis of the SDSS $i$-band images. Column 4: Effective radius quoted by T09, based on \citet{blanton05}. Columns 5, 6, \&\ 7: Values of the \sersic\ parameter $n$, the axis ratio $b/a$, and the total $I$ mag, all from the fits to the LRIS images.}
\end{deluxetable}

\begin{figure*}[!p]
\epsscale{1.0}
\plotone{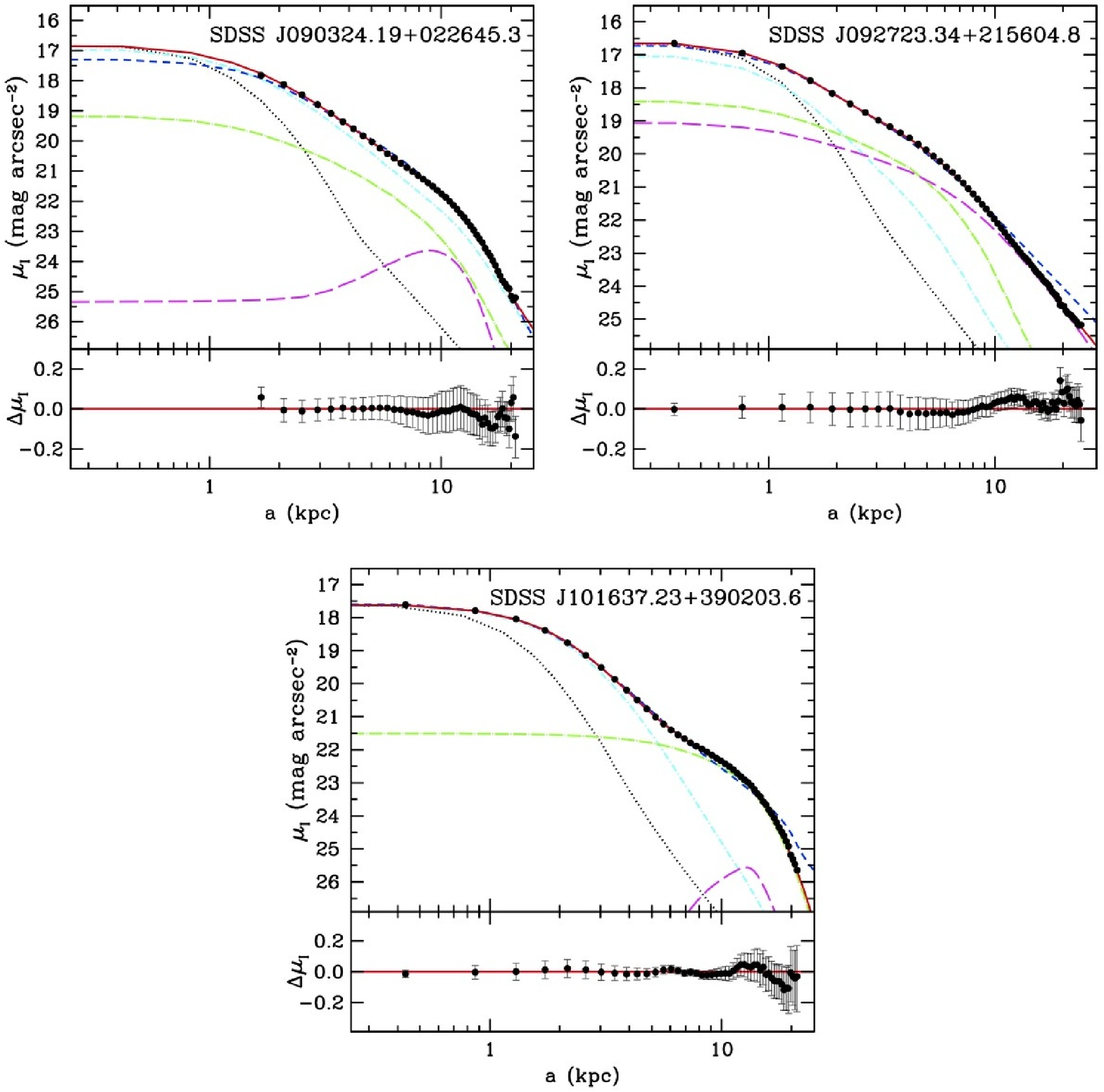}
\caption{Surface brightness profiles of all three targets. Data are shown as black dots, the multiple \sersic\ component fits are shown in red solid lines, and the single \sersic\ fits are shown in blue dashed lines. Individual components are plotted, from most compact to least compact, in cyan (dash-dot), green (long dash-dot), and magenta (long dash). The black dotted profiles show the PSFs scaled to the model centers. The difference between the best fit (multi-component) model and the data are shown under each comparison plot. Error bars are omitted in the surface-brightness plots for clarity but are included in the difference plots. Note that the component indicated by the magenta (long dash) component in SDSSJ0903 is decentered from the other 2 components to model the asymmetry in this galaxy, resulting in the non-monotonic profile. The magenta (long dash) component in SDSSJ1016 is a small, possibly interacting, companion galaxy projected on the outer part of the main galaxy profile.}
\label{ellipse}
\end{figure*}

We traced the surface brightness (SB) profile of our targets using the {\it ellipse} task in STSDAS. The radial-surface-brightness profiles along the major axes of the models and data, calculated from elliptical annuli sampling, are shown in Fig.~\ref{ellipse}. The PAs of the elliptical annuli were obtained from the single component \sersic\ fits and were held fixed, along with the center of each ellipse, for all measurements. The ellipticity, however, was allowed to vary, particularly since the fits of necessity include convolution with the PSF. The multi-component models typically remained well within 0.1 mag ($\frac{\Delta flux}{flux_{data}} < 0.1$) of the data out to at least 15 kpc. The single-component fits are slightly worse, but mostly still within 0.2 mag of the data. Note that, although the multi-component fits appear to be very close to the single-component fits in the 1-D plots, adding more components significantly reduced the $\chi^{2}$ in the 2-D fits. More importantly, systematic residuals, although still present, are greatly reduced in the multicomponent fits. The residuals from the single-component fits are shown as insets in the right-hand panels of Fig.~\ref{fits}.

\section{What is the Relation between Luminous Compact Passive Galaxies at High and Low Redshifts?}

Our single \sersic\ fits returned $R_{e}$ comparable to values in T09 for SDSSJ0903 and SDSSJ0927. However, for SDSSJ1016, which had the smallest listed $R_{e}$ of their sample, our fits gave a $R_{e}$ that was almost three times as large as theirs. Our SB plots show that the single \sersic\ profile fits the SDSSJ1016 data very well up to at least 6 kpc, so any significant decrease in the effective radius would certainly result in a much worse fit. We believe that this discrepancy is mostly resolved by our {\sc galfit} measurement of the SDSS $i$-band image (which we carried out at the suggestion of the referee). This determination gives a value of $R_e$ twice that quoted by T09 from the \citet{blanton05} catalog, which apparently is in error for this galaxy. The remaining difference from our LRIS value may be due to the lower resolution and depth of the SDSS images. In Fig. \ref{fits} we can see that SDSSJ1016 has an outer extended component that that appears to have faded into noise in the shallower SDSS exposures, possibly accounting for the increased $R_{e}$ that we measured. 

Single-component fits are fine for rough estimates of the galaxy morphologies and for comparison with other single-component fits, but we find that they leave rather large systematic residuals for all of our objects. By using more components for fitting we were able to significantly reduce these residuals. For example, from the surface-brightness plot of SDSSJ1016, it is especially clear that the model deviates from the data beyond $\sim 7$ kpc. Adding a second component improved the fit to the outer regions of the galaxy. In these multi-component models, each galaxy has one component with $R_{e} < 1$ kpc comprising from 32\%\ to 74\%\ of the total light, with the remainder distributed in much more extended, low-surface-brightness components. Note that, although we have used 3 components in our models for SDSSJ0903 and SDSSJ0927, the offset 3rd component in SDSSJ0903 is only present to model the asymmetry in the extended component of this galaxy, accounts for only $\sim3$\%\ of the total flux, and probably should not be considered to be a discrete physical entity. On the other hand, SDSSJ0927 appears to be a truly complex object, given that the 3 fitted components are nearly equally balanced in flux, and that the 2 outer components have very different ellipticities. Nevertheless, it is still the case that all three galaxies fit the pattern of a compact ($R_e<1$ kpc) core with a distinct more extended outer structure.

Both the single-component fits and the more complex structures of these low-redshift galaxies suggest that they are not as compact as most of those found at $z\sim2.5$. While they may have compact cores, they also have relatively extensive non-negligible wings that predominate in the outer regions. Wings having a similar surface-brightness relation to the cores would be visible within the dynamic range (typically $\sim5$ mag; see, e.g., \citealt{sto08}) accessible for at least some of the high-redshift galaxies. It is possible that these low-redshift objects are less massive compact galaxies that have acquired outer extended structures via some process such as accretion through minor mergers. SDSSJ1016 appears to have a small companion at the lower left of the data image in Fig.~\ref{fits}, which we have had to include in our model to avoid distorting the fit to the main galaxy. If this smaller object is gravitationally bound to this system, it may be an example of this minor merging scenario in action. At least some massive compact galaxies found at intermediate redshifts also have a similar compact core + extended-component configuration on a smaller scale. The two $z \sim 0.5$ objects studied by \citet{stockton10} each have a compact core with $R_{e} \lesssim 250$ pc containing $\gtrsim 30\%$ of the light, and another larger component with $R_{e} \sim 1.25$ kpc (note that the original paper quotes semimajor-axis values for $R_{e}$, whereas we give circularized values here). T09 also estimated the average ages and metallicities of the stellar populations of the galaxies in their sample, using the H$\beta$ and MgFe features in the Sloan spectra. They found that, compared with a carefully selected control sample of galaxies with similar masses and environments, but more typical values for $R_e$, the compact galaxies had populations with significantly younger ages ($\sim2$ Gyr vs. $\sim14$ Gyr) and somewhat higher metallicities ([Z/H]$\sim0.2$ vs. slightly less than solar). We have used a complementary approach of fitting Charlot-Bruzual spectral synthesis models (S. Charlot 2007, private communication\footnotemark)\footnotetext{These models are updated versions of those given by \citet{bruzual03}, the main improvement being the incorporation of thermally pulsating asymptotic-giant-branch stars into the models.} to the Sloan photometry. We have used models with a Chabrier initial mass function, and we have explored metallicities of 0.4, 1.0, and 2.5 times solar, and after confirming that models with large amounts of extinction gave poor fits to the photometry, we modeled variations in $A_V$ from 0.0 to 1.0, assuming a \citet{cal00} extinction law. We also tested both instantaneous bursts (i.e., forming all of the stars simultaneously) and more realistic models with exponentially declining star formation rates with e-folding times $\tau$ of up to 1.0 Gyr.  For solar metallicity models, the model ages were 4.25, 3.75, and 4.0 Gyr with $\tau = 0.5$, 0.4, and 0.5 Gyr for SDSSJ0903, SDSSJ0927, and SDSSJ1016, respectively, all with $A_V = 0$. These solar metallicity models were the best fits for the first two, but a 2.5 times solar instantaneous burst model with an age of 1.68 Gyr and $A_V = 0$ gave a significantly better fit for SDSSJ1016. The stellar masses we obtain from our best-fit models (in units of $10^{11}$ $M_{\odot}$) are 1.7, 1.5, and 0.8, to be compared with values of 1.28, 1.28, and 1.08 given by T09.
We thus have general agreement on both the fact that these galaxies are relatively young and on the stellar masses, and even some indication that some of the metallicities may be supersolar. These relatively young ages imply that the bulk of the stellar populations in these galaxies was formed at quite modest redshifts, ranging from $\sim0.4$ to $\sim0.8$.

Since all 3 of our targets appear to have larger diffuse components in addition to a more compact one, we suspect that many of the other galaxies in the T09 sample will also turn out to have these faint outer components. Deeper images of other objects in the T09 sample would easily determine whether most of them have structures as complex as the ones we have observed. These, as well as the 2 higher-redshift objects from \citet{stockton10}, appear already to have undergone some evolution and are no longer purely compact objects. This result continues to suggest that truly small, massive, and compact galaxies similar to those common at high redshifts are extremely rare in the low redshift universe. In addition, the best examples that have been found so far and studied in some detail seem not to be survivors from the population found at $z > 2$, but are instead fairly young objects, in the sense that the bulk of their stellar populations did not form at high redshifts. As emphasized by \citet{taylor10}, the scarcity of truly old massive compact galaxies in the low redshift universe cannot be explained by highly stochastic processes, such as major mergers. The driving mechanism behind the size evolution of the high-redshift population must operate in such a way that essentially all of them inevitably become incorporated into the sorts of galaxies observed in the local universe. Furthermore, the relatively young ages of both the T09 sample and the massive compact galaxies we have studied at $z\sim0.5$ \citep{stockton10} indicate that this compactness is a stage that is, in cosmological terms, a rather brief one.

We can now attempt a broad-brush picture of some of the relevant points regarding these compact massive galaxies, along with some of the crucial unanswered questions:

\begin{enumerate}
\item Their compactness and implied mass densities indicate that they must have formed via strongly dissipative processes, most likely involving mergers of gas-rich systems \citep{khochfar06,hopkins10,wuyts10,ricciardelli10}.
\item The very first massive galaxies seem to have formed in this way, given that almost all massive galaxies at $z\gtrsim2$ that do not show continuing star formation are quite compact.
\item The comparative youth of the stellar populations in the T09 sample and in the two \citet{stockton10} galaxies suggest that these galaxies formed via mechanisms similar to those at high redshifts, but at later epochs.  It would be of considerable interest to determine the mass fraction of very old ($\gtrsim10$ Gyr) stars in these systems.
\item The compact massive galaxies found so far at low and intermediate redshifts all have significant extended components that are absent (at similar scaled surface brightnesses) from at least some of the high redshift examples. There are at least two possibilities:
\begin{enumerate}
\item After formation, the compact galaxies may acquire these components over timescales of $\sim2$ billion years or so. This process may involve minor mergers or gas inflows, but, if so, the merger or inflow rate must be quite high. This conclusion follows from the observation that the absence or extreme rarity of ``naked'' compact massive galaxies at low redshift means that the extended components must be added via a fairly continuous rain of material rather than via a small number of discrete events. In addition, the total extended component mass involved in the cases studied so far ranges from several $\times10^{10}$ $M_{\odot}$ to over $10^{11}$ $M_{\odot}$, again suggesting a large amount of added material.
\item The extended components could be the expected distribution of pre-existing stars in the galaxies whose merger precipitated the strong dissipation required to form the compact cores of these galaxies \citep[e.g.,][and references therein]{wuyts10}. If so, we might expect that the stars in the these extended components might be somewhat older and of lower metallicity than the stars in the cores. Note that if the extended structures are actually disks that formed from the infall of new gas {\it after} the formation of the compact cores, the stellar populations would be younger than those of the cores. It would therefore be especially useful to obtain resolved spectroscopy of some of these galaxies to evaluate the nature of the stellar populations in the compact cores relative to those in the extended components.
\end{enumerate}
\end{enumerate}

Finally, our observation of SDSSJ1016 indicates the need to be cautious regarding some of the $R_e$ values listed in the SDSS DR6 NYU Value-Added Galaxy Catalog \citep*{blanton05}. While this catalog may be a good source for drawing up a list of candidate compact galaxies at low redshifts, it is clearly worthwhile to go back to the original SDSS images to check these candidates. Our initial skepticism about the feasibility of estimating $R_e$ values from the relatively low-resolution SDSS images has proved to be too pessimistic, at least for galaxies with (single-component) $R_e\sim1$ kpc at $z\sim0.2$. Nevertheless, even at $z\sim0.2$, imaging observations with greater depth and finer pixel scale are useful for exploring the detailed structure of such galaxies, which is not apparent from the SDSS images.

\acknowledgements 
We thank the anonymous referee for a careful reading of the paper and for making a number of detailed suggestions that greatly improved both its substance and its presentation. 
Funding for the Sloan Digital Sky Survey (SDSS) has been provided by the Alfred P. Sloan Foundation, the Participating Institutions, the National Aeronautics and Space Administration, the National Science Foundation, the U.S. Department of Energy, the Japanese Monbukagakusho, and the Max Planck Society. The SDSS Web site is http://www.sdss.org/.
The SDSS is managed by the Astrophysical Research Consortium (ARC) for the Participating Institutions. The Participating Institutions are The University of Chicago, Fermilab, the Institute for Advanced Study, the Japan Participation Group, The Johns Hopkins University, the Korean Scientist Group, Los Alamos National Laboratory, the Max-Planck-Institute for Astronomy (MPIA), the Max-Planck-Institute for Astrophysics (MPA), New Mexico State University, University of Pittsburgh, University of Portsmouth, Princeton University, the United States Naval Observatory, and the University of Washington.

 \end{document}